\documentclass{elsart}

\usepackage{natbib,epsfig}

\usepackage{amssymb}

\begin{document}

\def\ga{\lower.5ex\hbox{$\buildrel>\over\sim$}}
\def\la{\lower.5ex\hbox{$\buildrel<\over\sim$}}

\begin{frontmatter}



\title{Black hole mass estimates\\ from soft X-ray spectra}


\author[label1,label2]{Roberto Soria\thanksref{label3}}
\address[label1]{Harvard-Smithsonian CfA, 60 Garden st, 
  Cambridge, MA 02138, USA}
\address[label2]{Mullard Space Science Lab (UCL), 
  Holmbury St Mary, Surrey RH5 6NT, UK}
\thanks[label3]{Supported by a Marie Curie Outgoing International
  Fellowship}
\author[label4]{Zdenka Kuncic\thanksref{label5}}
\address[label4]{School of Physics, University of Sydney, NSW 2006, 
  Australia}
\thanks[label5]{Supported by a University of Sydney Bridging Support 
Research Grant}

\begin{abstract}
In the absence of direct kinematic measurements, the mass 
of an accreting black hole is sometimes inferred from the 
X-ray spectral parameters of its accretion disk; specifically, 
from the temperature and normalization of a disk-blackbody 
model fit. Suitable corrections have to be introduced 
when the accretion rate approaches or exceeds the Eddington 
limit. We summarize phenomenological models that can  
explain the very high state, with apparently higher disk 
temperatures and lower inner-disk radii. Conversely, 
ultraluminous X-ray sources often contain cooler disks 
with large characteristic radii. We introduce another 
phenomenological model for this accretion state. We argue 
that a standard disk dominates the radiative output 
for radii larger than a characteristic transition radius 
$R_{\rm c} \sim \dot{m} \times R_{\rm ISCO}$, where  
$\dot{m}$ is the accretion rate in Eddington units and 
$R_{\rm ISCO}$ is the innermost stable orbit. For 
$R_{\rm ISCO} < R < R_{\rm c}$, most of the accretion power 
is released via non-thermal processes. 
We predict the location of such sources in a luminosity-temperature 
plot. We conclude that BHs with masses $\sim 50$--$100 M_{\odot}$ 
accreting at $\dot{m} \sim 10$--$20$ may 
explain the X-ray properties of many ULXs.
\end{abstract}


\end{frontmatter}

% main text
\section{Standard disk in a thermal dominant state}
\label{}
Determining the masses of the accreting black holes (BHs) 
in ultraluminous X-ray sources (ULXs) is a key 
unsolved problem in X-ray astrophysics. In the absence 
of direct kinematic measurements of their mass functions, 
indirect methods based on X-ray spectral modelling have 
sometimes been used, by analogy with Galactic BH X-ray 
binaries, whose spectra can be approximated, 
in the ``canonical'' $0.3$--$10$ keV range, 
with a thermal component 
plus a power-law. The thermal component is consistent 
with disk-blackbody emission from an optically 
thick disk, while the power-law is generally attributed 
to inverse-Compton scattering of softer disk photons off 
high-energy electrons. The peak temperature and luminosity 
of the thermal component are good indicators 
of the size of the X-ray-emitting inner-disk region, 
which in turn is related to the BH mass. 

%In the simplest and most commonly used disk-blackbody 
%approximation \citep{mak86}, implemented as {\tt diskbb} 
%in {\small XSPEC} (Arnaud, 1996),
In the standard disk-blackbody 
approximation (Makishima et al., 1986, 2000), 
\begin{eqnarray}
L_{\rm disk} &=& 4 \pi \sigma (T_{\rm in}/\kappa)^4 
     (R_{\rm in}/\xi)^2 \equiv 4 \pi \sigma T_{\rm in}^4 r_{\rm in}^2\\
R_{\rm in} & =& R_{\rm ISCO} \equiv 6\alpha GM/c^2,
\end{eqnarray}
where: $L_{\rm disk}$ is the integrated (bolometric) 
disk luminosity; $T_{\rm in}$ is the peak colour temperature, 
which can be directly inferred from X-ray spectral 
fitting\footnote{the most commonly used implementation 
of the disk-blackbody model is {\tt diskbb} in {\small XSPEC} 
(Arnaud, 1996).}; 
$T_{\rm in}/\kappa$ is the peak effective temperature;  
$\kappa$ is the hardening factor ($1.5 \la \kappa \la 2.6$);  
$R_{\rm in}$ is the true inner-disk radius, assumed to coincide 
with the innermost stable circular orbit $R_{\rm ISCO}$; 
$r_{\rm in}$ is the apparent 
inner radius obtained from spectral fitting with the 
disk-blackbody approximation; $\alpha$ depends on the spin 
of the BH ($\alpha = 1$ for a Schwarzschild BH, $\alpha = 1/6$ 
for an extreme Kerr BH). The numerical factor $\xi$ 
was introduced (Kubota et al., 1998) to obtain a correctly 
normalized bolometric disk luminosity, taking into account that 
the fitted peak temperature occurs at $R=(49/36)R_{\rm in}$ 
because of the no-torque boundary conditions.  
With this approximation, and with the choice of $\kappa = 1.7$ 
(Shimura \& Takahara, 1995), the physical inner-disk radius 
is related to the fitted value by: 
\begin{equation}
R_{\rm in} \equiv (\xi^{1/2} \kappa)^2 r_{\rm in} 
      \approx (1.09)^2 r_{\rm in}. 
\end{equation} 
An alternative estimate of $\xi \kappa^2$, which takes 
into account relativistic corrections, suggests that 
\begin{equation}
R_{\rm in} \approx (1.35)^2 r_{\rm in}
\end{equation} 
(Fabian et al.,~2004), which 
is equivalent to a choice of $\kappa \approx 2.1$ for the hardening 
factor in (3). 
A hardening factor as high as $\kappa \approx 2.6$ was  
empirically estimated for some 
Galactic BHs in the high/soft state (Shrader \& Titarchuk, 2003).
 
%is the ratio between the inner disk radius and the radius at which 
%the disk luminosity peaks, and $g(i)$ takes into account 
%relativistic and geometric corrections. Those terms can be  
%lumped into a numerical factor of order 1 (Fabian et al.,~2004); 
%the combined correction to the inferred mass is small 
%($f_0 \approx 1.35$) for the purpose of this discussion.
The bolometric luminosity can be derived from 
the observed flux: 
\begin{equation}
L_{\rm disk} = 2 \pi d^2 f_{\rm bol} (\cos i)^{-1}, 
\end{equation}
where $d$ is the source distance and $i$ the viewing angle 
(the disk flux being higher in the direction 
perpendicular to the disk plane). 
In addition, the disk luminosity can be directly related  
to the accretion rate, from general energy-conservation 
priciples. Ignoring relativistic corrections,
%\begin{equation}
%\sigma [T_{\rm eff}(R_{\rm in}/\xi)]^4 \equiv (T_{\rm in}/\kappa)^4 
%\approx \frac{3GM\dot{M}}{8 \pi (R_{\rm in}/\xi)^3}\,(1-\xi^{1/2}),
%\end{equation}
\begin{equation}
L_{\rm disk} = \frac{GM\dot{M}}{2R_{\rm in}} \approx \frac{1}{12 \alpha} 
     \dot{M} c^2 \equiv \eta \dot{M} c^2. 
\end{equation}

From (1), (2), and (6) we have:  
\begin{eqnarray}
%M &=& \frac{(\xi \kappa^2)}{6 \alpha G (2 \sigma)^{1/2}} 
%   \frac{d}{(\cos i)^{1/2}} f_{\rm disk}^{1/2} T_{\rm in}^{-2}
%\nonumber\\
%  &=& \frac{(\xi \kappa^2)}{6 \alpha G (4 \pi \sigma)^{1/2}} 
%   L_{\rm disk}^{1/2} T_{\rm in}^{-2}
%\nonumber\\
M  &\approx& \frac{c^2 \xi \kappa^2 \eta}{G(\sigma \pi)^{1/2}} \, 
  L_{\rm disk}^{1/2} \, T_{\rm in}^{-2}\nonumber\\
  &\approx& 10.0 \, \left(\frac{\eta}{0.1}\right)\,
    \left(\frac{\xi \kappa^2}{1.19}\right)\,
    \left(\frac{L_{\rm disk}}{5 \times 10^{38} 
    {\rm{~erg~s}}^{-1}}\right)^{1/2}
    \, \left(\frac{kT_{\rm in}}{1  
    {\rm{~keV}}}\right)^{-2} \, M_{\odot},
\end{eqnarray}
which can also be expressed as $L_{\rm disk} \sim M^2 T_{\rm in}^{4}$. 
This is the fundamental evolutionary track of an accretion 
disk in the thermal-dominant state. As their accretion rate 
varies, individual sources are expected to move along tracks 
of constant $M$ and $R_{\rm in}$. 
This has been observationally verified for Galactic BHs 
(e.g., Kubota \& Makishima, 2004; Miller et al.,~2004), at least until 
their disk luminosity approaches the classical Eddington limit, 
$L_{\rm Edd} \approx 1.3 \times 10^{38} (M/M_{\odot})$ erg s$^{-1}$.
Despite the various approximations, (7) works well 
(within a factor of $2$) when applied to the masses 
of Galactic BHs, assuming radiative efficiencies $\la 0.2$. 
Theoretical and observational arguments suggest that 
stellar-mass BHs are typically found in a thermal-dominant state when 
their luminosity and corresponding accretion rate are 
$\sim 0.1$--$1$ times the classical spherical 
Eddington limit.

\section{Spectral arguments for intermediate-mass BHs}
\subsection{Soft X-ray emission in ULXs} 
ULX X-ray spectra can also be modelled 
with a power-law plus a thermal component, much cooler 
than that observed from Galactic BHs --- typically, with colour 
temperatures $kT_{\rm in} \sim 0.12$--$0.20$ keV for the brightest 
sources (Miller et al.,~2004; Feng \& Kaaret, 2005; 
Stobbart et al.,~2006). If (7) is applied, with total 
$0.3$--$10$ keV luminosities $\approx 1$--$3 \times 10^{40}$ 
erg s$^{-1}$, the inferred masses are $\sim 1000$--$2000 M_{\odot}$, 
or a factor of two lower if we consider only the luminosity 
in the thermal component. This argument has been 
used in support of the intermediate-mass BH interpretation 
of ULXs (Miller et al.,~2004). 

However, it is unclear whether this argument can be directly 
applied to ULXs. There is still little direct evidence 
that the ``soft excess'' in ULXs is due to disk emission, 
partly because of our limited observing band. 
Various alternative interpretations have been suggested 
(e.g., Gon\c{c}alves \& Soria, 2006; Stobbart et al.,~2006; 
Ebisawa et al., 2003). For example, analogous soft-excesses 
in AGN, and particularly in narrow-line 
Seyfert 1 galaxies, can be explained as blurred, 
ionized absorption (mostly in the $\sim 0.5$--$2$ keV 
band) and reprocessing of the primary 
power-law-like spectrum in a fast outflow (Gierli\'{n}ski 
\& Done, 2004; Chevallier et al., 2006). 
Gon\c{c}alves \& Soria (2006) showed that a similar interpretation 
can be applied to at least some ULXs, whose X-ray spectra are very 
similar to those of narrow-line Seyferts. Downscattering of harder 
photons by a cooler outflow is another mechanism proposed 
to explain the soft excess (King \& Pounds, 2003; 
Laming \& Titarchuk, 2004).
Even if we accept that the soft excess is indeed thermal 
emission from a disk (as we shall do in the rest of this paper),
we need to keep in mind that some of the disk-blackbody 
scaling relations summarized in Section 1 have been 
confirmed observationally for the thermal dominant 
state (Remillard \& McClintock, 2006), but have to be 
modified when a consistent fraction of the accretion 
power is not directly radiated by the disk (Section 3). 
%We summarize 
%simple parametric corrections in Section 3, and illustrate 
%their effects they have on the BH mass estimate.

\subsection{Cool disks and power-law component}
Two key issues need to be addressed by any ULX spectral model. 
The first is that the X-ray spectra of most ULXs 
are dominated by a broad, power-law-like component, at least in the limited 
energy range covered by {\it Chandra} and {\it XMM-Newton}.
The thermal component represents only $\sim 10$--$30$ per cent 
of the $0.3$--$10$ keV emission (Stobbart et al.,~2006; 
Feng \& Kaaret, 2005; Winter et al., 2006).
Physically, this suggests that most of the accretion power 
is not directly radiated by the disk, but is efficiently transferred 
in other forms (mechanical, thermal or magnetic energy) 
to an upscattering medium, and then partly radiated 
with a non-thermal (power-law-like) spectrum.
In Galactic stellar-mass BHs, this happens more commonly when 
the system is in the low/hard state, accreting at less 
than a few percent of their Eddington rate. But for transient sources, 
it also happens briefly near the peak of their outbursts, 
when they are probably accreting above their Eddington rate 
(very high state).

The second issue to be addressed 
is why the disk appears so cold for its luminosity. 
Even when we take into account only the luminosity in the fitted 
thermal component, ULX disks can radiate up to $\approx 10^{40}$ 
erg s$^{-1}$ with a much lower peak temperature 
than that observed from Galactic BHs in their high state, 
even though the latter are an order of magnitude less luminous.
A standard, truncated disk, replaced in the inner region
by a radiatively-inefficient flow (Narayan \& Yi 1994), 
would produce a power-law-dominated spectrum with a cold disk 
component (low/hard state).
However, if ULXs are in the low/hard state at luminosities 
$\approx 10^{40}$ erg s$^{-1}$ (and hence, at mass accretion rates 
$\ga 10^6 M_{\odot}$ yr$^{-1}$), their BH masses 
should be $\ga 10^4 M_{\odot}$. 
The formation of such massive remnants in the local Universe 
is difficult to explain with existing models of stellar 
or star cluster evolution, 
and the existence of primordial intermediate-mass BHs is not 
yet supported by independent evidence.
Therefore, we need to consider alternative scenarios, 
in which the accretion rate is high, but the disk is or appears 
cooler than a standard disk in a thermal dominant state, 
for the same accretion rate and BH mass. 

In the rest of this paper, we will discuss phenomenological 
corrections to the disk-blackbody equations, to estimate 
what physical effects may produce more luminous but cooler disks.
We will also illustrate the effects that those corrections 
have on the BH mass estimates.

%This may 
%happen because the hotter part of it is not directly 
%visible, or because the inner disk directly radiates 
%only a small fraction of the accretion power, the rest 
%being extracted non-thermally (``chilled disk'' scenario). 
%One can build phenomenological models by introducing 
%simple parametric corrections to the disk-blackbody 
%set of equations.

\section{Parametrical modifications of the disk-blackbody equations}

\subsection{Constant removal of power}
The first plausible scenario we consider is that a constant 
fraction of power $\epsilon > 0$ is extracted from the disk 
at each radius (independent of radius); for example, 
to power a corona. 
%$\epsilon$ may be increasing with the accretion rate.  
The radiative flux from the disk becomes
\begin{equation}
\sigma T_{\rm eff}(R)^4 \approx (1-\epsilon) \, \frac{3GM\dot{M}}{8 \pi R^3}.
\end{equation}
As a result, the peak temperature 
$T_{\rm in} \rightarrow T_{\rm in}' = (1-\epsilon)^{1/4} T_{\rm in}$, and 
$L_{\rm disk} \rightarrow L_{\rm disk}' = (1-\epsilon) L_{\rm disk}$. 
Neither $R_{\rm in}$ nor $r_{\rm in}$ changes.
The emitted spectrum is still a disk blackbody. 
As $\epsilon$ increases, the disk becomes cooler and less 
luminous, for a fixed accretion rate, and its parameters move 
downwards along the same $L_{\rm disk} \sim M^2 T_{\rm in}^{4}$ track.
This scenario is equivalent to changing the effective 
accretion rate $\dot{M} \rightarrow \dot{M}' = (1-\epsilon) \dot{M}$.
It follows from the previous argument that the BH mass 
is still given by (7), independent of $\epsilon$.
Hence, removal of a constant fraction of accretion power 
does make the disk cooler but does not explain the high 
BH masses inferred from ULX spectral fitting.

\subsection{Hardening of the disk photons}
Another simple modification to the standard-disk scenario 
is to assume that an increasing fraction of the emitted disk 
photons are upscattered by relativistic electrons 
at the disk photosphere. This is equivalent to 
increasing the hardening factor $\kappa$, 
typically from $\approx 1.7$ to $\approx 2.5$ 
(Davis et al., 2005; 2006). 
Observationally, this spectral hardening occurs at  
accretion rates near or just above Eddington.

In the $(T_{\rm in}, L_{\rm disk})$ plane, the effect is to move 
the source horizontally to the right: higher colour 
temperature $T_{\rm in}$ but same effective temperature $T_{\rm in}/\kappa$
and same luminosity $L_{\rm disk}$. The true inner-disk radius 
$R_{\rm in}$ does not change, while the apparent disk radius decreases: 
$r_{\rm in} \rightarrow r_{\rm in}' = [\kappa/(\kappa + \Delta \kappa)]^{2} r_{\rm in}$.
This does not affect the mass estimate, if one has properly 
taken into account its dependence on $T_{\rm in}/\kappa$, 
as we did in (7).

\subsection{Partial covering of the disk}
The third effect that needs to be taken into account 
is a partial covering of the disk surface: a fraction $X$ of photons 
may be absorbed or more generally removed from the disk-blackbody 
spectrum by clouds or a moderately optically-thick corona. They 
are re-emitted in other spectral bands or components (e.g., 
X-ray power-law, or infrared). In a self-consistent analysis, 
one must of course account for all upscattering, downscattering 
and absorption effects, redistributing the total available 
accretion power into the various components. But for the purpose 
of relating the {\it disk} emission parameters to the BH mass, 
we can simply consider those photons as being lost from 
the disk spectrum. The effect is to reduce the observed disk 
flux and consequently the inferred luminosity: 
$L_{\rm disk} \rightarrow L_{\rm disk}' = (1-X) L_{\rm disk}$.
The spectral shape is not altered.
The peak colour temperature $T_{\rm in}$ is not changed. 
$R_{\rm in}$ stays the same but the apparent 
disk radius decreases: 
$r_{\rm in} \rightarrow r_{\rm in}' = (1-X)^{1/2} r_{\rm in}$, 
because $r_{\rm in}$ is indirectly derived from the flux 
normalization.
The BH mass inferred from the observed luminosity and 
temperature via (7) is a factor $(1-X)^{1/2}$ less than the true mass.
 
The combined effect of the modifications described in Sections 3.1, 3.2 
and 3.3 is to move the location of the source {\it to the right-hand-side 
of its thermal track}, in the $(T_{\rm in}, L_{\rm disk})$ plane (Figure 1). 
The precise displacement depends on the relative importance of 
the parameters $\epsilon, \Delta \kappa$ and $X$. 
Physically, it leads to lower estimates for the radius $r_{\rm in}$ 
(sometimes much lower than the innermost stable orbit), 
and may lead to an underestimate of the BH mass if not properly 
accounted for. These empirical corrections may be used 
to explain the {\it very high state (or steep-power-law state)} 
of accreting black holes (Remillard \& McClintock, 2006).
They do not help explain the high masses and large 
radii inferred for ULXs.

\section{A two-zone structure of the inflow}
For high accretion rates, it is plausible to speculate 
that accretion power may not be uniformely removed 
from the whole disk (Section 3.1) and/or disk photons may not 
be uniformly ``lost'' independent of radius. More likely, 
those losses may be strongly concentrated in the inner disk region 
(Done \& Kubota, 2006; Soria \& Kuncic, 2007b).
For example, the inner disk may be disrupted or simply occulted 
by an optically-thick outflow (King \& Pounds, 2003). 
Or, all of its photons may be comptonized by an optically-thick 
corona and emerge as a power-law or broken power-law 
component (Stobbart et al.,~2006).
Alternatively, the inner disk 
may still be present and visible, but most of its accretion 
power is extracted via non-radiative processes, 
such as mechanical outflows or Poynting flux (Kuncic \& Bicknell, 2004). 
Some of the power released through those channels is then 
converted to non-thermal photons.

Hence, a possible phenomenological way to correct 
the standard disk equations is to assume the presence of 
a transition radius $R_{\rm c} \gg R_{\rm ISCO}$.
For $R > R_{\rm c}$, the inflow can be approximated 
by a standard disk, such that
\begin{equation}
L_{\rm disk} \approx 4 \pi \sigma T_{\rm c}^4 R_{\rm c}^2,
\end{equation}
where $T_{\rm c} \equiv T(R_{\rm c})$ is the maximum 
(observable) disk temperature.
For $R_{\rm ISCO} < R < R_{\rm c}$, we assume that all the accretion power 
comes out in the power-law-like component, and the standard disk
either is not directly visible or gives a negligible direct 
contribution.

Regardless of the details, in this scenario, (2) no longer holds 
or is no longer relevant to the observed spectrum. It is replaced by 
\begin{equation}
R_{\rm c} = F R_{\rm ISCO}
\end{equation}
with $F = F(\dot{M}) > 1$. 
The visible disk radiates only the accretion power released 
from the outer radius to $R_{\rm c}$. The effective radiative 
efficiency of that part of the inflow 
$\sim M/R_{\rm c} \sim 1/(12\alpha F) < 0.1$. 
Therefore, we have (cf.~(6),(7)):
\begin{eqnarray}
L_{\rm disk} &\approx & \frac{GM\dot{M}}{2R_{\rm c}} 
     \approx \frac{1}{12 \alpha F} \dot{M} c^2 
     \equiv \frac{\eta}{F} \dot{M} c^2 \\
M  &\approx & \frac{10.0}{F} \left(\frac{\eta}{0.1}\right) 
    \left(\frac{\xi \kappa^2}{1.19}\right)\,
    \left(\frac{L_{\rm disk}}{5 \times 10^{38} 
    {\rm{~erg~s}}^{-1}}\right)^{1/2}
    \, \left(\frac{kT_{\rm in}}{1  
    {\rm{~keV}}}\right)^{-2} \, M_{\odot} \nonumber\\
   &\approx & \frac{790}{F}  \left(\frac{\eta}{0.1}\right) 
    \left(\frac{\xi \kappa^2}{1.19}\right)\,
    \left(\frac{L_{\rm disk}}{5 \times 10^{39} 
    {\rm{~erg~s}}^{-1}}\right)^{1/2}
    \, \left(\frac{kT_{\rm in}}{0.2  
    {\rm{~keV}}}\right)^{-2} \, M_{\odot},
\end{eqnarray}
where $L_{\rm disk} \sim 5 \times 10^{39}$ erg s$^{-1}$ and 
$kT_{\rm in} \sim 0.2$ keV are typical values found in ULXs. 
We speculate that the high masses inferred 
for some of those systems are not evidence 
of intermediate-mass BHs, 
but are simply a consequence of a high value $F \ga 10$ 
not properly taken into account in the disk modelling.  
Defining a dimensionless accretion rate 
\begin{equation}
\dot{m} \equiv \dot{M}/\dot{M}_{\rm Edd} \equiv \frac{0.1 \dot{M} c^2}{L_{\rm Edd}},
\end{equation}
the bolometric disk luminosity integrated for $R>R_{\rm c}$ can 
more usefully be expressed as
\begin{equation}
L_{\rm disk} \approx \left(\frac{\dot{m}}{F}\right) \, 
    \left(\frac{\eta}{0.1}\right) L_{\rm Edd}. 
\end{equation}

\begin{center}
\begin{figure}
\epsfig{figure=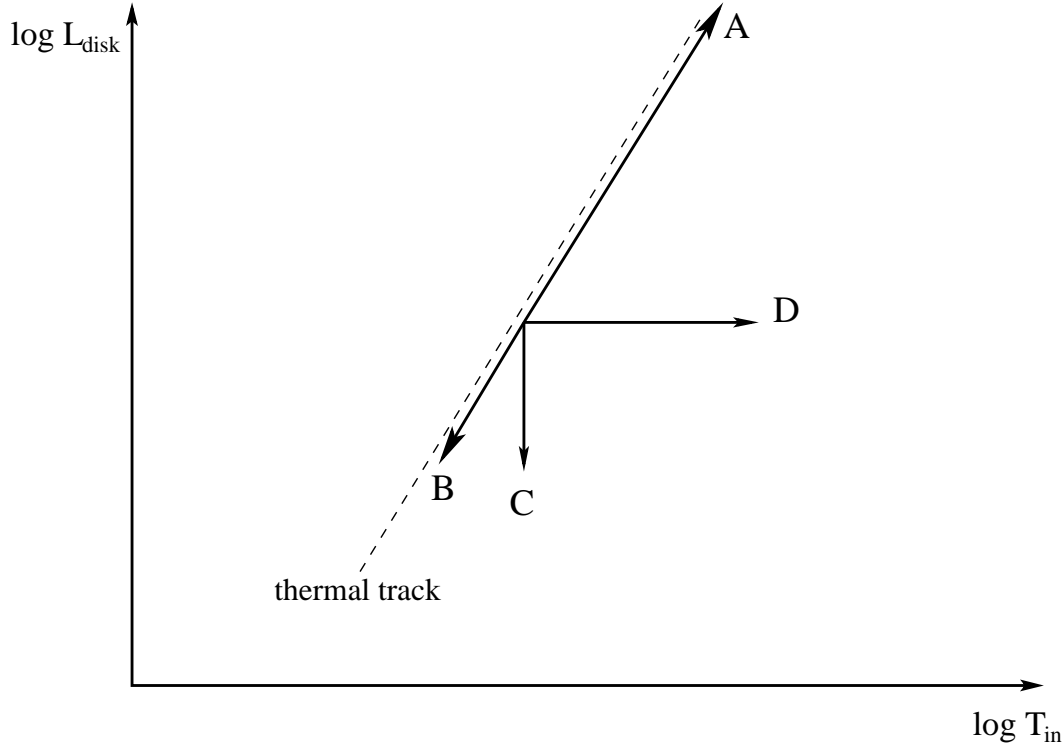, width=14cm, angle=0}
\caption{Evolution of the source parameters in 
the $(T_{\rm in}, L_{\rm disk})$ plane, due to various competing 
effects at high accretion rates. Dashed line: thermal track, 
characterized by $L_{\rm disk} \sim T_{\rm in}^4 M^2$;
A: effect of increasing the accretion rate; B: effect of injecting 
some of the accretion power into a corona or outflow 
(uniformly at all radii); C: effect of removing some 
of the disk photons via absorption or upscattering into 
the power-law component; D: effect of spectral hardening 
due to electron scattering near the disk photosphere 
(equivalent to increasing the colour temperature without 
changing the effective temperature).}
\end{figure}
\end{center}

\section{Observational predictions}

It was suggested that the disk structure parameterized in Section 4 
is applicable to sources with very high 
accretion rates, $\dot{m} > 1$ (Done \& Kubota, 2006; 
Soria, 2007; Soria \& Kuncic, 2007b).
To estimate the BH mass and accretion rate of a source in that state, 
from the observed disk luminosity and temperature, we need 
to estimate $F= R_{\rm c}/R_{\rm ISCO}$, either by measuring it directly, 
or by understanding how it depends on $\dot{m}$.
The simplest way to estimate $F$, at least as an order of magnitude,  
is to recall that in this scenario, all of the accretion power 
released at $R > R_{\rm c}$ is radiated by the disk, 
and some or most of the power released 
at $R_{\rm ISCO} < R < R_{\rm c}$ is eventually radiated 
as power-law photons, without contributing 
to the thermal disk component. 
Since the accretion power liberated down to a radius $R$ 
scales $\sim 1/R$, one can estimate that 
$F \ga L_{\rm tot}/L_{\rm disk}$.  
The inequality sign takes into account the fact 
that the radiative efficiency of the non-thermal processes 
in the inner region is less than 
the efficiency of blackbody emission in a standard disk.
For the most luminous nearby ULXs, 
extrapolating from the relative fluxes of the power-law 
and thermal components in the $0.3$--$10$ keV band, the 
luminosity ratio $L_{\rm tot}/L_{\rm disk} \sim 3$--$20$; in a few other, 
more distant ULXs, there is only a lower limit $\ga 10$ for this ratio.
Hence, we infer that $F \ga 10$ for typical ULXs.
 
From the fitted luminosity and temperature 
of the disk (9), typical values of 
$R_{\rm c} \ga 5000$ km are found for many ULXs, i.e., 
$\sim 100$ times larger than typical inner-disk radii 
in stellar-mass BH. If $F \ga 10$, it means that BH masses 
in those ULXs are only required to be $\la 10$ times larger 
than typical BH masses in Galactic systems. This simple 
argument suggests that ULX masses are $\la 100 M_{\odot}$. 
%bolometric disk luminosities $\la L_{\rm Edd}$ and total X-ray 
%luminosities $\la 3 \times L_{\rm Edd}$ (Soria \& Kuncic 2007b).

The location or evolutionary track of a source in 
the $(T_{\rm c}, L_{\rm disk})$ plane provides 
another observational constraint. To predict the displacement 
of a source from its thermal track when $F>1$, 
we need to understand how $F$ and $T_{\rm c}$ 
vary as a function of $\dot{m}$. For a fixed BH mass, 
the radiative flux equation of a standard disk tells us 
that $T_{\rm c} \sim R_{\rm c}^{-3/4} \dot{m}^{1/4}$.
It was suggested (Poutanen et al.,~2007; 
Begelman et al.,~2007; Soria \& Kuncic, 2007b) 
that $R_{\rm c} \sim \dot{m}$, based on plausible physical 
processes that may form such a transition radius. 
If so, from (9) we expect that, as the accretion rate $\dot{m}$ increases, 
a source will move along a track defined by $L_{\rm disk} \approx$ constant, 
$T_{\rm c} \sim \dot{m}^{-1/2}$ (Figure 2).

The luminosity track is very sensitive to the radial temperature distribution 
on the disk, at $R>R_{\rm c}$. Disk models with a distribution flatter 
than $R^{-3/4}$ are sometimes used. For example, it was found 
(Kubota et al.,~2005) that $T \sim R^{-0.7}$ may provide 
a more accurate fit to the X-ray spectral data 
of Galactic BHs. In that case, we expect
$L_{\rm disk} \sim T_{\rm c}^{-0.44}$, with 
$T_{\rm c} \sim \dot{m}^{-0.45}$. For a disk dominated 
by X-ray irradiation (as may be the case if $R \gg R_{\rm ISCO}$), 
$T \sim R^{-0.5}$. In that case, 
$L_{\rm disk} \sim T_{\rm c}^{-4}$, with 
$T_{\rm c} \sim \dot{m}^{-0.25}$. See Soria \& Kuncic (2007b) 
for further discussions and a comparison with observations.

Regardless of the precise functional forms of $F(\dot{m})$ 
and $T_{\rm c}(R_{\rm c})$, the significant result is 
that we expect a source to move {\it to the left-hand-side 
of its thermal track}, as the accretion rate increases (Soria 2007).
Along that track, the luminosity of the disk component 
stays constant or increases, even if the observed peak 
temperature decreases. 
We speculate that the effects described in Sections 3.1, 3.2 and 3.3 
dominate when $\dot{m} \ga 1$, and the shift 
to larger transition radii and lower peak temperatures 
becomes the dominant term for $\dot{m} \gg 1$. 
Thus, we predict that an accreting BH may initially move up along 
its thermal track, then slightly down and to the right, 
then cross the thermal track and move to the left, 
for increasing accretion rates
(Figure 2). As it does so, we also expect the source 
to become more and more dominated by non-thermal 
emission components (including Compton scattering). 
We also suggest that transient 
stellar-mass BHs are more likely to proceed in the opposite 
direction: moving from the low hard state 
directly to the ultraluminous state on the left-hand-side 
of the diagram, at the beginning 
of the outburst, when the accretion rate 
is highest; then they may move to the right, 
spend some time in the very high state and high/soft 
state as the accretion rate declines, and finally return 
to the low/hard state.

The fitted X-ray properties of many bright ULXs are qualitatively 
consistent with this phenomenological scenario. Their disks 
are very luminous but have a low peak temperature. The total 
power output is dominated by a power-law component, perhaps 
in addition to other non-thermal and non-radiative mechanisms 
(e.g., outflows). Their characteristic variability timescales 
are consistent with large disk radii (Soria \& Kuncic, 2007b). 
For example, let us consider an accreting BH with $M \approx 50 M_{\odot}$, 
and a disk luminosity $\approx 5 \times 10^{39}$ erg s$^{-1}$ 
$\approx L_{\rm Edd}$. If $\dot{m} \approx 1$ and $F =1$, 
we expect its spectrum to be dominated by the disk 
component, with a peak temperature $\approx 0.8$ keV 
and a characteristic size of the X-ray emitting region 
$\sim 6GM/c^2 \sim 500$ km. Although the luminosity 
is consistent with an average ULX, its spectral properties 
are not. However, let us allow for $\dot{m} \approx 20$, 
and consequently $F \approx 20$. In that case, we expect 
that the source will still have 
$L_{\rm disk} \approx L_{\rm Edd} \approx 5 \times 10^{39}$ 
erg s$^{-1}$, but a peak temperature 
$\approx 0.8 \times 20^{-0.5}$ keV $\approx 0.18$ keV, 
a characteristic radius $\sim 20 \times 6GM/c^2 \sim 10,000$ km, 
and a dominant X-ray power-law component with a luminosity 
$\ga 5 \times 10^{39}$ erg s$^{-1}$. These are more 
typical ULX parameters. If we had observed such a source, 
determining only its fitted values of $L_{\rm disk}$ and $T_{\rm c}$, 
and inserting those values directly into (7) 
instead (9), we would have incorrectly 
interpreted the source as an intermediate-mass BH with 
$M \approx 1000 M_{\odot}$.

\begin{center}
\begin{figure}
\epsfig{figure=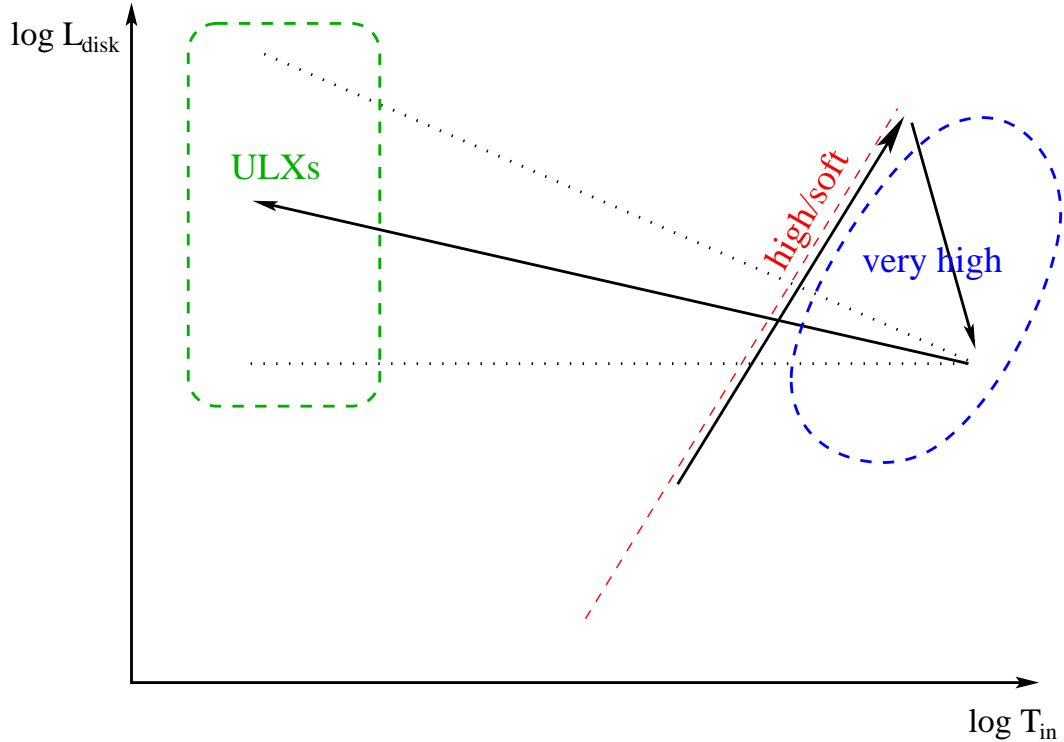, width=14cm, angle=0}
\caption{Suggested X-ray spectral evolution of BH accretion disks 
for accretion rates $0.1 \la \dot{m} \la 20$. Sources 
first move up the thermal track, until their disk luminosity 
$L_{\rm disk} \approx L_{\rm Edd}$ (high/soft state). Then, they 
move to the very high state (or steep-power-law state), generally 
characterized by a higher peak temperature, lower disk luminosity 
and a dominant, steep power-law component. Finally, as $\dot{m}$ increases 
further, they may move to the ULX region, characterized by 
a large transition radius between outer disk and inner non-thermal 
region, and a dominant, flatter power-law component. 
The location of the thermal track (high/soft state) is 
determined by the BH mass; the horizontal shift towards lower apparent 
temperatures is mostly determined by the accretion rate.}
\end{figure}
\end{center}

\section{Conclusions}
It is now recognized that it may not be possible to estimate 
BH masses in ULXs using the standard disk-blackbody model. 
If we accept that the soft excess is the disk 
emission component (which at this stage is only one possible 
hypothesis), we need to account for the fact that ULX spectra 
are power-law dominated and their thermal component has 
a remarkably low peak temperature. 
We briefly reviewed simple phenomenological corrections 
to the standard disk model, which have been used to explain 
the very high state of Galactic BHs. However, 
they do not help explaining the ULX spectral properties.
We then discussed the effects of introducing 
a transition radius $R_{\rm c} \gg 6GM/c^2$, 
such that for $R> R_{\rm c}$ the accretion flow is a standard
disk, while for $R< R_{\rm c}$ the disk is either 
drained of all its power via non-thermal 
processes, or its photons are completely reprocessed 
by an optically-thick comptonizing medium.
In this case, the peak disk emission comes from 
$R \approx R_{\rm c}$, while the inner part of the inflow 
emits with a power-law-like spectrum.

We speculate that $R_{\rm c}/R_{\rm ISCO} \sim \dot{m}$, 
where $\dot{m}$ is the accretion rate in Eddington units.
With this parameterization, for a fixed BH mass, the peak 
colour temperature $T_{\rm c} \sim \dot{m}^p$ with $p \approx -0.5$, 
and $L_{\rm disk} \sim T_{\rm c}^{q}$, with $q \le 0$.
The predicted effect is to move the source 
to the left-hand-side of its thermal track 
in the $(T_{\rm c}, L_{\rm disk})$ plane.
This behaviour may be incorrectly read as standard disk emission 
from a more massive BH. We argue that this had lead 
to the flawed intermediate-mass BH interpretation.
In our scenario, if the accretion rate $\dot{m} \sim 20$, 
the BH mass can be $\sim 50$--$100 M_{\odot}$; hence, perhaps 
still consistent with their formation via stellar processes, 
in very special circumstances. On the other hand, 
the BH mass may not be much smaller than $\sim 50 M_{\odot}$, 
if we assume that $L_{\rm disk} \la L_{\rm Edd}$ 
and $L_{\rm tot} \la$ a few $L_{\rm Edd}$. In this scenario,
the peak flux contribution from the disk may come 
from $\sim 100$ gravitational radii. The dependence 
of the transition radius on the accretion rate 
needs to be taken into account also when scaling 
the frequency of X-ray quasi-periodic oscillations 
between sources of different mass. Lower frequencies 
may be caused by either an increase of $\dot{m}$ 
or an increase of $M$.

In summary, we speculate that ULXs may be defined as accreting BHs 
with masses
a few times higher than Galactic BHs, and persistent accretion 
rates at least an order of magnitude above their Eddington limit.
None of them has been seen to move along its thermal track, 
typical of accretion rates $\sim 0.1$--$1$ Eddington. Perhaps 
this is due to the nature of their massive Roche-lobe-filling 
donor stars, which provide a consistently higher mass transfer 
rate. A physical interpretation of the transition radius, 
as well as a more detailed comparison with observed X-ray spectral 
and timing properties of some Galactic BHs and ULXs will be presented 
elsewhere (Soria \& Kuncic, 2007b).

\section*{Acknowledgements}
We thank G. V. Bicknell, A. C. Gon\c{c}alves and J. E. McClintock 
for useful discussions and suggestions.

%a Schwarzschild BH (Schwarzschild, 1916).

% \appendix

% \section{}
% \label{}

% Bibliographic references with the natbib package:
% Parenthetical: \citep{Bai92} produces (Bailyn 1992).
% Textual: \citet{Bai95} produces Bailyn et al. (1995).
% An affix and part of a reference:
%   \citep[e.g.][Ch. 2]{Bar76}
%   produces (e.g. Barnes et al. 1976, Ch. 2).

\end{document}